\documentclass[10pt,aps,prx,twocolumn,superscriptaddress,nofootinbib]{revtex4-2}
\usepackage[utf8]{inputenc}
\usepackage[T1]{fontenc}
\usepackage{braket}
\usepackage{amsfonts, mathtools}
\usepackage{dcolumn}
\usepackage{booktabs}
\usepackage{siunitx}
\usepackage[breaklinks=true,colorlinks,citecolor=blue,linkcolor=blue,urlcolor=blue]{hyperref}
\usepackage{comment}
\usepackage{orcidlink}
\usepackage[ruled,vlined]{algorithm2e}
\usepackage{xcolor}
\usepackage[normalem]{ulem}
\usepackage{float}
\usepackage{tabularx}
\usepackage{multirow}

\usepackage{listings}
\lstdefinestyle{kipu}{
  basicstyle=\ttfamily\small,
  numbers=left,
  numberstyle=\tiny,
  stepnumber=1,
  numbersep=8pt,
  showstringspaces=false,
  breaklines=true,
  frame=single,
  rulecolor=\color{black!20},
  tabsize=2,
  captionpos=b,
  upquote=true
}
\lstdefinelanguage{json}{
  basicstyle=\ttfamily\small,
  literate=
   *{0}{{{\color{black}0}}}{1}
    {1}{{{\color{black}1}}}{1}
    {2}{{{\color{black}2}}}{1}
    {3}{{{\color{black}3}}}{1}
    {4}{{{\color{black}4}}}{1}
    {5}{{{\color{black}5}}}{1}
    {6}{{{\color{black}6}}}{1}
    {7}{{{\color{black}7}}}{1}
    {8}{{{\color{black}8}}}{1}
    {9}{{{\color{black}9}}}{1}
    {:}{{{\color{black}{:}}}}{1}
    {,}{{{\color{black}{,}}}}{1}
    {"}{{{\color{black}{"}}}}{1},
}

\usepackage{soul}

\begin{document}

\title{Quantum-enhanced satellite image classification}

\author{Qi Zhang$^{\orcidlink{0000-0001-6223-5516}}$}
\email{qi.zhang@kipu-quantum.com}
\affiliation{Kipu Quantum, Greifswalderstrasse 212, 10405 Berlin, Germany}

\author{Anton Simen$^{\orcidlink{0000-0001-8863-4806}}$}
\affiliation{Kipu Quantum, Greifswalderstrasse 212, 10405 Berlin, Germany}
\affiliation{Department of Physical Chemistry, University of the Basque Country UPV/EHU, Apartado 644, 48080 Bilbao, Spain}

\author{Carlos Flores-Garrig\'os$^{\orcidlink{0009-0000-9735-5411}}$}
\affiliation{Kipu Quantum, Greifswalderstrasse 212, 10405 Berlin, Germany}
\affiliation{IDAL, Electronic Engineering Department, ETSE-UV, University of Valencia, Avgda. Universitat s/n, 46100 Burjassot, Valencia, Spain}

\author{Gabriel Alvarado Barrios$^{\orcidlink{0000-0002-8684-4209}}$}
\affiliation{Kipu Quantum, Greifswalderstrasse 212, 10405 Berlin, Germany}

\author{Paolo A. Erdman}
\affiliation{Kipu Quantum, Greifswalderstrasse 212, 10405 Berlin, Germany}

\author{Enrique Solano$^{\orcidlink{0000-0002-8602-1181}}$}
\affiliation{Kipu Quantum, Greifswalderstrasse 212, 10405 Berlin, Germany}

\author{Aaron C. Kemp$^{\orcidlink{0009-0000-6313-0417}}$}
\email{aaronkemp@kpmg.com}
\affiliation{KPMG LLP, 2 Manhattan West, New York, NY 10001}

\author{Vincent Beltrani$^{\orcidlink{0009-0004-9854-4459}}$}
\affiliation{IBM T. J. Watson Research Center, 1101 Kitchawan Rd. Yorktown Heights, NY 10598}

\author{Vedangi Pathak$^{\orcidlink{0000-0001-6335-6539}}$}
\affiliation{IBM T. J. Watson Research Center, 1101 Kitchawan Rd. Yorktown Heights, NY 10598}

\author{Hamed Mohammadbagherpoor$^{\orcidlink{0009-0006-1482-9970}}$}
\email{hamed.mohammadbagherpoor@ibm.com}
\affiliation{IBM T. J. Watson Research Center, 1101 Kitchawan Rd. Yorktown Heights, NY 10598}


\begin{abstract}
We demonstrate the application of a quantum feature extraction method to enhance multi-class image classification for space applications. By harnessing the dynamics of many-body spin Hamiltonians, the method generates expressive quantum features that, when combined with classical processing, lead to quantum-enhanced classification accuracy. 
Using a strong and well-established ResNet50 baseline, we achieved a maximum classical accuracy of 83\%, which can be improved to 84\% with a transfer learning approach. In contrast, applying our quantum-classical method the performance is increased to 87\% accuracy, demonstrating a clear and reproducible improvement over robust classical approaches.
Implemented on several of IBM’s quantum processors, our hybrid quantum-classical approach delivers consistent gains of 2–3\% in absolute accuracy.
These results highlight the practical potential of current and near-term quantum processors in high-stakes, data-driven domains such as satellite imaging and remote sensing, while suggesting broader applicability in real-world machine learning tasks.
\end{abstract}

\maketitle

\section{Introduction}\label{sec:introduction}
Quantum technologies are poised to become a transformative toolset across multiple industrial domains, including optimization \cite{farhi2014quantumopt, abbas2024challengesopt, chandarana2025runtimeopt, hegade2022digitizedopt, simen2025branchopt}, machine learning \cite{biamonte2017quantum, caro2021generalization, havlivcek2019supervised}, secure communications \cite{pirandola2020advances}, and high-precision sensing \cite{degen2017quantum}. The space sector, still an emerging but highly promising area for quantum innovation, stands to benefit significantly from these advances, providing new computational, communication, and sensing capabilities that meaningfully complement and extend classical approaches.

Space applications are inherently constrained by complex operational environments and rapidly expanding data volumes, particularly in Earth observation, large satellite constellations, and autonomous mission planning \cite{treedata, tehsin2023satellite, rashid2021comprehensivecrop}. Quantum computing, quantum sensing, quantum communication, and quantum machine learning collectively offer promising avenues to overcome these limitations, addressing computational bottlenecks, enhancing measurement precision, enabling more secure space-to-ground information flows, and improving the accuracy and expressiveness of machine-learning models applied to space data \cite{belenchia2022quantum,dequal2025quantum}.
Quantum optimization can support more efficient mission planning, satellite routing, and debris-removal strategies by addressing complex combinatorial problems. Quantum communication enables secure space-to-ground links through technologies such as satellite-based QKD, while quantum sensing promises higher-precision navigation and Earth-observation measurements for climate monitoring, resource mapping, and scientific missions.

Among quantum technologies, quantum machine learning (QML) is suitable for space applications because it can construct expressive data representations. Quantum feature maps encode classical inputs into high-dimensional Hilbert spaces, allowing classical algorithms to operate on quantum-induced embeddings~\cite{nkilloran2019, havlivcek2019supervised, biamonte2017qml, mengoni2019kernel}. Previous work have been demonstrated that quantum systems can extract complex information from images, beyond classical reach \cite{simen2024digital, simen2025digitized, simen2025quenchedquantumfeaturemaps}. Quantum dynamics that are classically intractable \cite{visuri2025annedcqo, king2023quantum, king2025beyond} are good candidates to obtain information that can enhance classical machine learning models \cite{cimini2025largereservoirboson, kornjavca2024largereservoirneutral, ciceri2025enhancedibmhsbc, bokov2026luqpiprivileged}. This trend will accelerate with future large satellite constellations. QML can improve feature extraction and classification for tasks such as land-use monitoring, environmental modeling, anomaly detection, and other remote-sensing applications. These advances can benefit sectors ranging from agriculture and insurance to urban planning, climate resilience, and biodiversity management.

\begin{figure*}[!t]
    \centering
     \includegraphics[width=\textwidth]{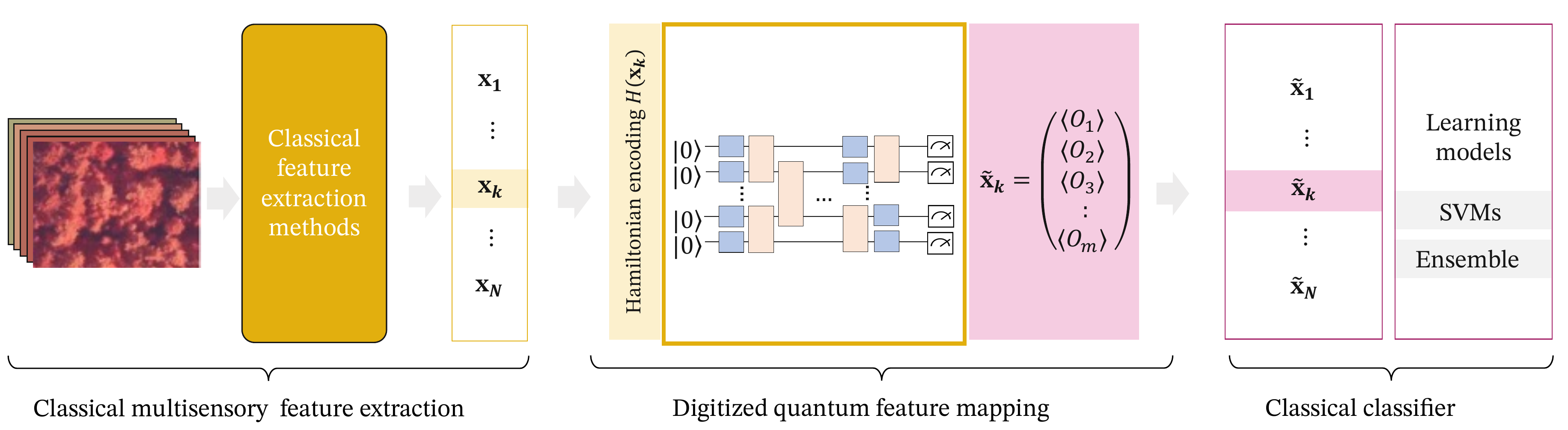}
    \caption{Overview of the quantum-enhanced classification pipeline. Multisensor
    remote-sensing imagery is first mapped to an n-dimensional classical
    feature vector (in this case, n = 15, 120 and 156). These features parametrize an n-qubit Hamiltonian,
    which is processed through Digitized Quantum Feature extraction (DQFE)
    using a counterdiabatic evolution protocol. Local measurements yield
    one-body expectation values and two-body correlation functions, which
    are then used as input to a classical classifier such as gradient
    boosting or random forests.}
    \label{fig:sketch}
\end{figure*}

In this work, we address multi-class image classification using multi-sensor Earth-observation data. By leveraging a Hamiltonian-based quantum feature-extraction approach, known as Digitized Quantum Feature Extraction (DQFE) \cite{simen2025digitized}, using counterdiabatic (CD) protocols \cite{del2013shortcutsadolfo} in the impulse regime \cite{hegade2022dcqo, cadavid2024efficientimpulse, Cadavid_2024, Cadavid_2025}, our model extracts features contained in the distribution of low-energy states as well as in the non-adiabatic transitions for given Hamiltonians that encode the feature vectors. Our experiments are conducted on the TreeSatAI benchmark, a real-world remote-sensing dataset that integrates three complementary sensing modalities: Sentinel-1 SAR data, Sentinel-2 multispectral imagery, and high-resolution aerial photographs, covering 15 tree-genus classes. To ensure compatibility with near-term quantum hardware, we construct a balanced and challenging 5-class subset following established reduction protocols.

Beyond a single reduction strategy, we explore four different feature-reduction settings, projecting the data to 15, 120, and 156 features, respectively. For each setting, a dedicated ResNet-50 architecture is trained, with its final layer adapted to produce embeddings of the corresponding dimensionality. This design allows us to target different quantum hardware backends and embedding capacities, specifically IBM AER (Simulator), IBM BOSTON and IBM PITTSBURGH (Heron r3), and IBM KINGSTON (Heron r2).
Across all configurations and hardware platforms, we consistently observe improved classification performance when combining classical features with quantum-enhanced features generated via DQFE, compared to purely classical pipelines. Among the classical baselines, the best performance is achieved with the 120-feature reduction, reaching an accuracy of approximately 84 percent. When this representation is further processed using DQFE and executed on IBM BOSTON hardware, the accuracy increases to approximately 86.5 percent. These results demonstrate a robust and reproducible quantum enhancement across multiple reduction strategies, hardware backends, and validation runs, highlighting the potential of hybrid classical–quantum feature representations in realistic remote-sensing applications.

These results illustrate that quantum feature extraction can provide value even with today’s noisy, near-term devices, and they highlight the growing potential of quantum machine learning for operational space applications.

\section{Methodology}
\label{sec:methodology}

Our quantum-enhanced image classification pipeline consists of three main stages, sketched in Fig.~\ref{fig:sketch}:
(i)~classical image feature extraction, (ii) Hamiltonian-based quantum
feature generation using Digitized Quantum Feature extraction (DQFE), and
(iii) classical classification on quantum-derived features.

This procedure for quantum feature extraction, targeting both industrial and academic applications, can be implemented using superconducting quantum circuits on IBM Quantum systems. A commercially available implementation of the DQFE-based quantum-enhanced feature extraction service is offered by Rimay \cite{rimay}.

\subsection{Image feature extraction}

To ensure compatibility with current quantum hardware, the input data must
be mapped to a feature space whose dimensionality does not exceed the
available number of qubits. In our case, this constraint requires reducing
the original image data, potentially high-dimensional and multi-channel, to a compact tabular representation of size \( n \leq 156 \). We therefore
consider reduction scenarios with \( n \in \{15, 90, 120, 156\} \), chosen
to match the qubit capacities of the targeted quantum processors.

To perform this dimensionality reduction, we employ a pretrained
ResNet-50 model as a feature extractor. All convolutional layers are
frozen, and a fully connected layer with \( n \) neurons is appended,
followed by an output layer with five neurons corresponding to the
five tree-genus classes. Only the newly added dense layers are trained,
using a validation set to monitor performance. After training, the final
classification layer is removed, and the network is truncated at the
penultimate layer, yielding an \( n \)-dimensional feature vector for
each image. This procedure is applied to both training and test sets,
resulting in a tabular dataset \( X \in \mathbb{R}^{N \times n} \),
where \( N \) denotes the number of samples and \( n \leq 156 \).

\subsection{Digitized quantum feature extraction}

For each sample, the n-dimensional feature vector
$\mathbf{x} = (x_1, \ldots, x_{n})$ is encoded into a quantum circuit by means of a digitized counterdiabatic (CD) evolution protocol. In particular, the feature vector is encoded into a spin-glass Hamiltonian $H_F(\mathbf{x})$ acting on $n$ qubits \cite{simen2025quenchedquantumfeaturemaps, simen2024digital, simen2025digitized}. The Hamiltonian consists of two terms: \textit{one-body terms}, where each feature $x_i$ modulates a
          local operator acting on qubit $i$ and encodes sample-specific
          information; and \textit{two-body interaction terms}, whose strengths are derived from the mutual information between pairs of features \cite{simen2025quenchedquantumfeaturemaps, simen2025digitized}, thereby incorporating classical feature correlations into the interaction graph.

To generate quantum features, the system is initialized in the ground state of the transverse-field Hamiltonian $H_i = \sum_{i=1}^{n} \sigma^x_i$ and then undergoes a discretized counterdiabatic (CD) evolution in the impulse regime \cite{Cadavid_2024, Cadavid_2025} that connects the initial transverse-field Hamiltonian $H_i$ to the instance-specific Hamiltonian $H_F(\mathbf{x})$. The resulting quantum state encodes both the local structure of the classical features and the higher order dependencies introduced by the two-body interaction terms. Since each image in the dataset corresponds to a distinct Hamiltonian $H_F(x)$, a separate quantum circuit is constructed and executed for every sample. The image-specific, spin-glass Hamiltonian, $H_F(x)$, is therefore given by
\begin{equation}
    H_F(\mathbf{x}) = \sum_{i=1}x_i\sigma^z_i + \sum_{(i,j)\in G}m_{ij}\sigma^z_i\sigma^z_j,
\end{equation}
where $x_i \in \textbf{x}$ and $m_{ij} = I(x_i, x_j)$ is the mutual information between the feature $i$ and feature $j$, where the couplings are selected from some interaction graph $G$. The interaction graph is given by a permutation of variables in the dataset in order to maximise the interaction strenghts in $G$ \cite{simen2025digitized}. A single-step counterdiabatic evolution in the impulse regime is performed in order to quench the encoding Hamiltonian. The final state is a complex misture of low-energy states and other states resulting from non-adiabatic excitations. 

Therefore, for each feature vector in the dataset we compute a quantum circuit with the method described above. For each circuit several measurement shots are performed, in order to have statistically reliable estimates of the relevant observables. From the resulting measurement outcomes we extract the quantum-enhanced feature representation, given by
\begin{equation}
    \tilde{\mathbf{x}} = \left(
    \langle\mathcal{O}_1\rangle,
    \langle\mathcal{O}_2\rangle, ...,
    \langle\mathcal{O}_m\rangle
    \right),
\end{equation}
where the estimated expectations, $\langle\mathcal{O}_i\rangle$,  are given by $n$ one-body expectation values $\langle \sigma^z_i \rangle$ and the two-body correlation functions $\langle \sigma^z_i \sigma^z_{j} \rangle$ for arbitrary pairs $(i,j)$, used for the downstream classical classification.



\subsection{Classification}

\begin{figure}[!tb]
    \centering
    \includegraphics[width=\linewidth]{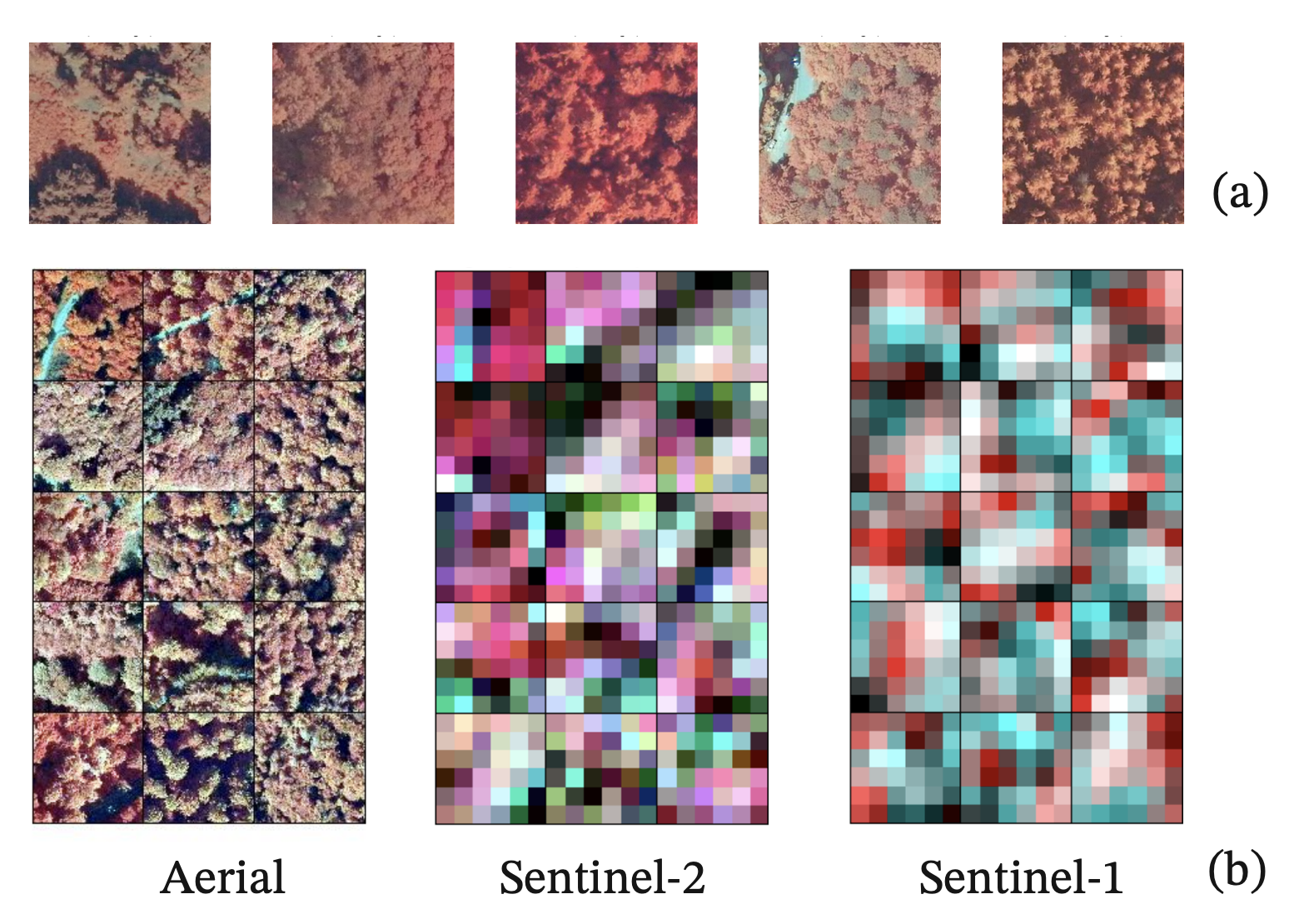}
    \caption{Overview of the reduced TreeSatAI dataset. {(a)}~Representative aerial patches from the five selected
    classes used in the hardware-compatible benchmark.
    {(b)} Example of the three sensing modalities available for each
    ground location: aerial imagery, Sentinel\textendash2 multispectral
    measurements, and Sentinel\textendash1 radar data. These heterogeneous
    inputs are reduced to $n$ classical features before quantum processing.}
    \label{fig:treesatai}
\end{figure}

The final step is performed entirely classically. We train ensemble-based
models such as gradient boosting and random forests, which are well suited
to tabular feature sets. The classifiers operate either on the
quantum-derived features alone or on the combined set of classical and
quantum features. Identical classical hyperparameters are used across
experiments to isolate the effect of quantum feature generation.

\section{Results}

\subsection{Image dataset preparation}

We address multi-class image classification in remote sensing imagery using the TreeSatAI dataset \cite{treedata}, a multisensor benchmark comprising of image triplets of high-resolution aerial photographs, Sentinel-2 multispectral imagery, and Sentinel-1 synthetic aperture radar (SAR) data (see Fig. \ref{fig:treesatai} for representative samples). The complete dataset contains 15 tree-genus classes. However, executing the DQFE workflow on physical quantum hardware necessitates limiting the total number of circuit evaluations. To ensure hardware feasibility, we construct a reduced benchmark by selecting the five most challenging classes based on structural similarity metrics. Each class comprises 200 training samples and 40 test samples, yielding a balanced dataset of 1,000 training samples and 200 test samples across all five classes.

We apply the full hybrid classical--quantum pipeline described in Sec.~\ref{sec:methodology}. To accommodate different quantum backends, we consider multiple feature-reduction scenarios in which the dimensionality of the classical feature representation is constrained to at most the number of available qubits on the target hardware. In particular, we report representative results obtained on four IBM quantum backends: IBM\_AER using 15 qubits, IBM\_BOSTON and IBM\_PITTSBURGH both using 120 qubits (from 156 available), and IBM\_KINGSTON utilizing all 156 qubits. The choice of 120 features for IBM\_BOSTON and IBM\_PITTSBURGH was initially selected to match the qubit count of alternative hardware architectures under consideration; this dimensionality also proved beneficial for classical baseline performance. This strategy enables a direct one-to-one embedding of classical features into the quantum circuit for all considered devices.

Across all hardware configurations, we compare three learning paradigms:
(i) a purely classical model based on ResNet-derived features,
(ii) a quantum-only model using features generated exclusively via DQFE,
and (iii) a hybrid classical--quantum model that combines both feature
representations. 
In all cases, classification is performed using a Random Forest (RF) classifier within a unified training pipeline that includes feature scaling and model selection. Hyperparameter optimization follows an identical procedure across all models: a grid search over the same parameter space using 5-fold cross-validation with 10 repetitions. While the optimization strategy remains consistent, the resulting optimal hyperparameters may differ between classical, quantum, and hybrid feature spaces, reflecting the distinct characteristics of each representation. To assess statistical robustness, all models with their selected hyperparameters are subsequently evaluated across multiple random training seeds, confirming that the observed performance differences are stable and not driven by specific data splits or initializations.

As shown in table \ref{tab:hardware_results}, the strongest purely classical baseline is obtained with a 120-dimensional ResNet-based representation, achieving an accuracy of approximately 84\%. When quantum feature extraction is introduced via classical simulation on IBM\_AER (using 15 qubits), the hybrid classical--quantum approach already demonstrates improved performance, consistently outperforming the corresponding purely classical baseline and validating the potential of DQFE-generated features in an idealized, noise-free setting.

Moving to physical quantum hardware, comparable performance gains are observed on IBM\_KINGSTON (156 qubits), where the hybrid models maintain their advantage over classical counterparts. Similarly, on IBM\_BOSTON (using 120 qubits), the hybrid model reaches an accuracy of approximately 86.5\%, representing a clear improvement over both the classical and quantum-only baselines. These results indicate that DQFE-generated quantum features can enhance classification performance across different hardware scales and topologies.
Notably, in the majority of evaluated scenarios, the hybrid classical--quantum approach yields the best overall performance. However, an exception is observed on IBM\_PITTSBURGH (using 120 qubits), where the purely quantum model achieves the highest accuracy of approximately 87.0\%, slightly surpassing the corresponding hybrid configuration. This result suggests that, for specific hardware topologies and noise characteristics, the quantum feature representation alone may already capture highly discriminative information sufficient for the task at hand, without requiring complementary classical features.

\begin{table}[h]
\centering
\renewcommand{\arraystretch}{1.2}
\begin{tabular}{|c|l|c|c|}
\hline
Dataset & Backend/Model & Var./Qubits & Accuracy (\%) \\
\hline
\multirow[c]{4}{*}{Hybrid} 
    & AER (sim.) & 15  & 83.5 \\
    & KINGSTON   & 156 & 81.5 \\
    & BOSTON     & 120 & \textbf{86.5} \\
    & PITTSBURG  & 120 & \textbf{86.5} \\
\hline
\multirow[c]{4}{*}{Quantum} 
    & AER (sim.) & 15  & 82.0 \\
    & KINGSTON   & 156 & 81.0 \\
    & BOSTON     & 120 & 86.0 \\
    & PITTSBURG  & 120 & \textbf{87.0} \\
\hline
\multirow[c]{7}{*}{Classical}
    & ResNet50       & 15  & 82.5 \\
    & ResNet50       & 90  & 81.5 \\
    & ResNet50       & 120 & 83.0 \\
    & ResNet50       & 156 & 79.5 \\
    & ResNet50 + RF  & 15  & 81.0 \\
    & ResNet50 + RF  & 156 & 79.5 \\
    & ResNet50 + RF  & 120 & \textbf{84.0} \\
\hline
\end{tabular}
\vspace{6pt}
\caption{Classification accuracy (\%) for classical, quantum-only, and hybrid classical-quantum models across different IBM quantum hardware backends. For each backend, the dimensionality of the classical feature representation is matched to the number of available qubits. In the classical segment, the number of variables are given by the number of nodes in the last hidden layer before the output one. The best result for each segment is highlighted in bold.}
\label{tab:hardware_results}
\end{table}

\begin{figure*}[!tb]
    \centering
    \includegraphics[width=0.85\textwidth]{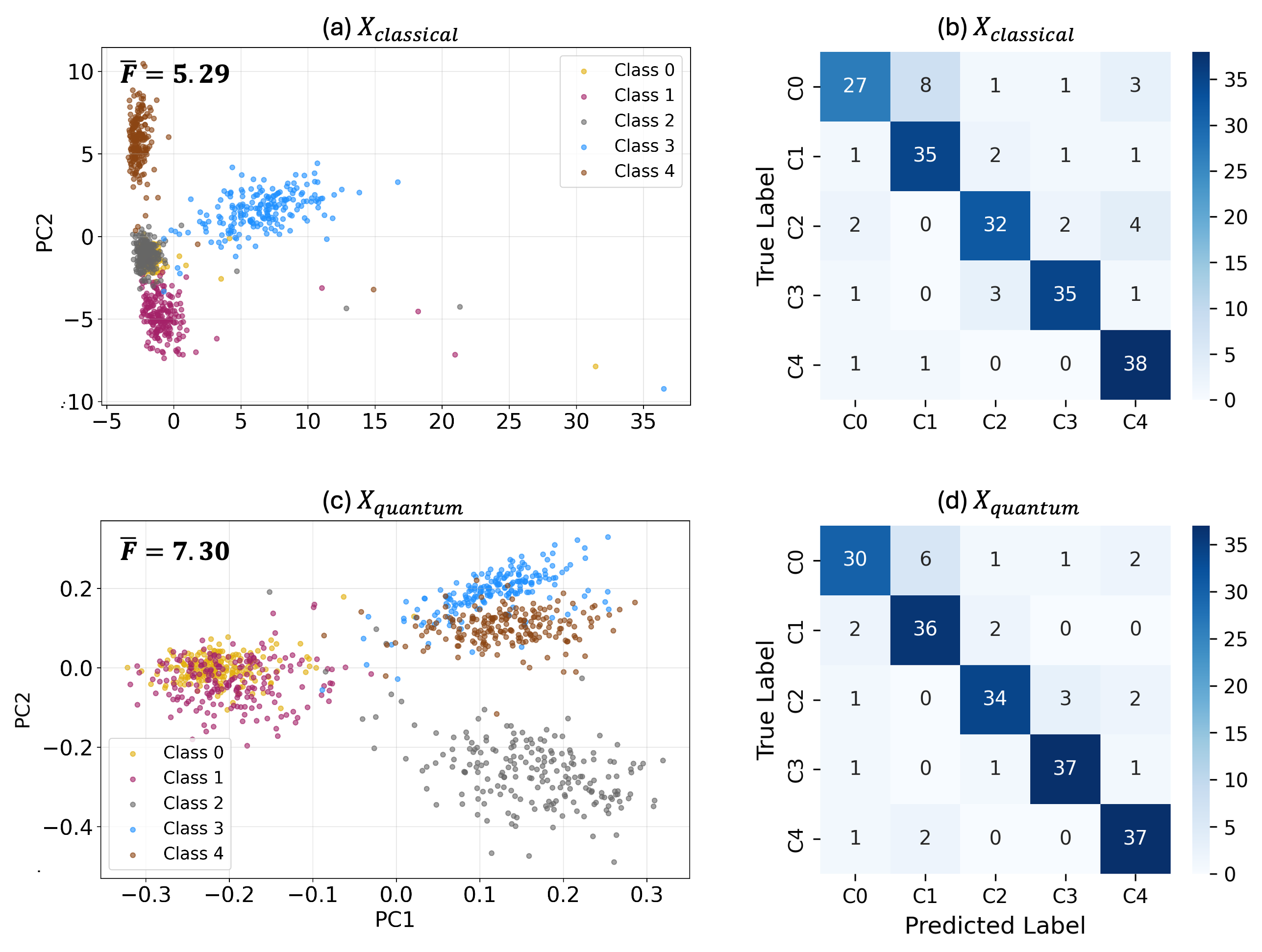}
    \caption{Two-dimensional PCA projections and their mean Fisher information, $\bar{F}$, of the (a) classical and (c) quantum feature representations and the respective classification performances on classical and quantum datasets for the TreeSatAI reduced benchmark.}
    \label{fig:pca_features}
\end{figure*}

A key observation from these results is the remarkable consistency of quantum enhancement: DQFE provides a 2--3\% absolute accuracy improvement across all quantum backends, independent of the classical baseline performance. This holds true whether the classical baseline is relatively modest (79.5\% with 156 features) or already strong (84\%  with 120 features). This consistent lift demonstrates that quantum features are not simply bridging a performance gap in scenarios with poor classical representations, but rather providing additive discriminative power even when classical methods already have strong performance. The exception on IBM\_PITTSBURGH, where the quantum-only model reaches 87\%, further underscores the value of quantum features under specific hardware conditions. These results provide compelling evidence that the observed gains originate from the quantum feature-generation mechanism itself, supporting a reproducible advantage that persists across different baseline conditions and hardware configurations in this remote-sensing classification task.

To gain some insight into the discriminative structure of the feature representations, we analyze two-dimensional projections of both classical and quantum feature spaces using the first two principal components variables derived from the PCA analysis
(PC1 and PC2), as shown in Fig.~\ref{fig:pca_features} (a) and (c). While both the classical ResNet-derived features and the DQFE-generated quantum features exhibit clear clustering for four out of the five classes, notable differences arise in the degree of class overlap. In the classical feature space, two classes show a near-complete overlap, making them difficult to distinguish. In contrast, although two classes also overlap in the quantum feature space, the separation between them is more pronounced, indicating an improved, though not perfect, discriminability after quantum feature extraction.

The remaining ambiguity is primarily associated with Classes~0 and~1, which correspond to two \emph{Pinus} genera and therefore share strong spectral and structural similarities. This increased difficulty is
consistent with domain knowledge and is reflected in both classical and
quantum representations. Notably, however, the overlap between these two
classes is substantially reduced in the quantum feature space, indicating
that the DQFE transformation enhances class-discriminative information
along directions that are not fully captured by the classical features.

This behavior is further corroborated by the confusion matrix obtained for
the IBM\_PITTSBURGH backend, shown in
Fig.~\ref{fig:pca_features}. The matrix reveals that the
majority of misclassifications occur between Classes~0 and~1, while the
remaining classes are classified with high accuracy. The strong diagonal
structure and limited off-diagonal entries provide additional evidence
that the improved separability observed in the PCA projections translates
directly into enhanced classification performance, particularly for the
purely quantum model on the IBM\_PITTSBURGH device.

\section{Discussion and conclusion}
\label{sec:conclusion}

This work demonstrates that quantum feature extraction via the DQFE
workflow leads to consistent and reproducible performance improvements
for multi-class image classification. While the strongest purely classical
models achieve accuracies between 79.5\% and 84\%, depending on the
feature dimensionality, incorporating quantum-derived features, either in
hybrid classical—quantum configurations or in purely quantum feature
settings, consistently yields absolute accuracy improvements of
approximately 2-3\%. In the majority of evaluated scenarios, the hybrid
approach achieves the best overall performance, reaching 81.5\% on IBM\_KINGSTON, 83.5\% on
IBM\_AER and 86.5\% on IBM\_BOSTON, thereby surpassing their respective
classical baselines.

A notable exception is observed on IBM\_PITTSBURGH, where the purely
quantum feature model achieves the highest accuracy of 87\%, slightly outperforming the corresponding hybrid configuration. This suggests that under specific hardware topologies and noise characteristics, quantum feature representations alone can capture sufficient discriminative information for the classification task. In this context, IBM's new Nighthawk architecture represents a promising future direction, as its modified hardware topology would in principle enable our algorithm to embed more information into the quantum protocol. As this architecture matures and throughput improves, it may offer substantial advantages for DQFE-based feature generation.

Analysis of the feature spaces through PCA projections reveals that DQFE-generated quantum features enhance class separability compared to classical ResNet-derived features, particularly for structurally similar classes such as the two \textit{Pinus} genera. While both representations exhibit clustering for four out of five classes, the quantum feature space demonstrates more pronounced separation between overlapping classes, with this improved discriminability translating directly into enhanced classification performance as evidenced by confusion matrix analysis.

These results provide compelling evidence that the observed gains originate from the quantum feature-generation mechanism itself, supporting a reproducible advantage that persists across different baseline conditions and hardware configurations. The DQFE workflow thus represents a viable approach for leveraging today's and near-term quantum devices to enhance classical machine learning pipelines, establishing a pathway to demonstrate practical quantum utility beyond purely theoretical capability in high-impact commercial domains, extending to applications such as satellite imagery and aerial photography classification, remote sensing, climate analytics, risk modeling, and related data-intensive fields.

\twocolumngrid
\bibliography{bibfile}

\end{document}